# Practical PV energy harvesting under real indoor lighting conditions

Bastien Politi[a,b], Stéphanie Parola[a], Antoine Gademer[a,c], Diane Pegart[a], Marie Piquemil[b], Alain Foucaran[a], Nicolas Camara[a,c]

[a] IES, University of Montpellier, CNRS, Montpellier, France
[b] Bureaux A Partager SAS, Paris, France
[c] EPF Graduate School of Engineering, Montpellier, France

ABSTRACT

Indoor light can be used as a new energy source to power µW low consumption wireless sensor networks (WSNs), but for wireless electronic devices consuming tens of mW, it is still challenging. The challenge comes from the low level of irradiance and from the several kinds of source combinations varying in time (multi-spectral direct, reflective, and scattered mix of artificial and natural light). This article describes a simple and reliable method that provides a model-based evaluation of the harvestable energy from any real indoor light environment. This method uses 'real condition' indoor light spectral measurements with a spectrometer as well as 'controlled condition' optoelectrical characteristics of the photovoltaic solar cells. The model-based evaluation of the harvestable energy has been compared with real microsource prototypes based on commercial photovoltaic cells powering commercial wireless e-ink display (more than 10 mW consumption averaged on a day). In this article, we show that it is possible to evaluate the harvestable energy, for several days of indoor light exposure, with an error lower than 6 %. Our method, with such an accuracy range, will be a helpful tool to assist engineers and researchers in designing light energy harvesting systems and more generally could find wide application in the growing IoT ecosystem.

## 1. Introduction

The number of low-power wireless devices is increasing significantly, and as a result, there is more and more research focusing on new ways to supply energy to these devices (Mathúna et al., 2008). Indeed, the growth in energy demand induced by this new field of application poses unprecedented challenges to provide the energy needed to operate these devices (Mathews et al., 2019). It also requires a review of how this energy is supplied and produced while minimizing its economic and environmental impact. One practical approach investigated for this purpose is harvesting light energy from the surroundings of the device. Past research has proven the viability of this approach in outdoor environments (Shaikh and Zeadally, 2016). In an indoor environment, where radiated levels are low, light energy harvesting has been identified as an effective method to provide enough power to low-power electronic systems such as wireless sensor networks (Matiko et al., 2014). Moreover, harvesting energy from light has demonstrated its capability as a means to achieve battery-free applications (Brunelli et al., 2009; Wang et al., 2016).

However, when it comes to considering energy harvesting for indoor applications, the difficulty in characterizing the harvestable power becomes substantial. Even though Standard Test Conditions (STC) from IEC 60904-3 are not entirely representative of the behavior of photovoltaic (PV) converters in real-life conditions, it is still a practical way to compare them. As for indoor conditions, the situation is somewhat different. The existing standards ISO 8995:2002 and CIE S 008/E define the level of luminosity required depending on the task associated with a workplace. The unit of this appropriate luminosity level is given in a semi-empirical unit of measure called lux. The lux (or lumen per m²) value $E_v$ is given by the following equation:

$$E_v = K_{cd} \int_0^\infty I_\lambda \, Y_\lambda \, d\lambda , \qquad (1)$$

**Nomenclature**

| | | | |
|---|---|---|---|
| a-Si:H | hydrogenated amorphous silicon | LED | Light-Emitting Diode |
| CFL | compact fluorescent light | Li-Po | Lithium-Polymer |
| $\eta^{SQ}$ | theoretical SQ solar cell efficiency | MAPE | mean absolute percentage error |
| $\eta^{meas}$ | measured solar cell efficiency | MPPT | maximum power point tracking |
| $E_g$ | bandgap energy | $n$ | diode's ideality factor |
| EQE | external quantum efficiency | pc-Si | polycrystalline silicon |
| $E_v$ | illuminance | PCE | power conversion efficiency (PCE) |
| $FF^{SQ}$ | theoretical fill factor from SQ model | $P_{in}$ | incident power |
| $FF^{meas}$ | measured fill factor | $\Phi_{p,\lambda}$ | incident photon flow |
| FOCV | fractional open-circuit voltage | $P^{meas}(V)$ | measured power -voltage |
| GaAs | gallium arsenide | $P^{meas}_{max}$ | measured maximum power |
| $h$ | Planck's constant | $P^{mod}(V)$ | calculated power -voltage |
| $I_\lambda$ | irradiance | $P^{mod}_{max}$ | calculated maximum power |
| $J^{SQ}_{SC}$ | short circuit SQ current density | PMIC | power management integrated circuit |
| $J^{QE}_{SC}$ | calculated short circuit current density | $R_s$ | parasitic series resistance |
| J-V | current density-voltage | $R_{sh}$ | parasitic shunt resistance |
| $J^{SQ}_0$ | theoretical SQ saturation current density | SMU | Source Meter Unit |
| $J^{meas}_{dark}(V)$ | Measured dark current density- voltage | SQ | related to Shockley-Quiesser model |
| $J_{ph}$ | photocurrent density | STC | Standard Test Conditions |
| $J_{SC}$ | short-circuit current density | $T$ | photovoltaic cell temperature |
| $J^{mod}(V)$ | estimated current density-voltage (model) | $V^{SQ}_{OC}$ | theoretical SQ open-circuit voltage |
| $J^{meas}(V)$. | measured current density-voltage | WSN | wireless sensor network |
| $k$ | Boltzmann's constant | $Y_\lambda$ | photopic luminosity function |
| $K_{cd}$ | maximum spectral efficiency coefficient | | |

where $Y_\lambda$ is a function related to the human eye's cones sensitivity to light (shown in Fig. 1) also called the photopic luminosity function, $I_\lambda$ is the irradiance and $K_{cd}$ is a coefficient called maximum spectral efficiency, defined as 686 lux.W$^{-1}$.m$^{-2}$ in such a way that irradiance of 1000 W.m$^{-2}$ in standard condition (AM 1.5G) corresponds to 100 klux.

Nevertheless, when considering how much energy could be harvested from PV converters, the first issue arises: the luminosity in lux is not a reliable quantification of the incident harvestable power (Wang et al., 2010). Indeed, as an example, Fig. 1 presents the emission spectra from 3 different light sources: i) the natural light through a window, ii) a compact fluorescent light (CFL), and iii) a light-emitting diode (LED). The three spectra have been measured with a commercial calibrated spectroradiometer, the StellarRAD from *StellarNet Inc.,* equipped with a CR2 cosine receptor with a wavelength range from 350 nm to 1100 nm, adapted to indoor environments. Spectra are set to obtain a 1000 lux illuminance level, giving three different levels of irradiance for each light source: 7 W.m$^{-2}$ for natural light through the window, 3 W.m$^{-2}$ for the CFL source, and 4 W.m$^{-2}$ for the LED source.

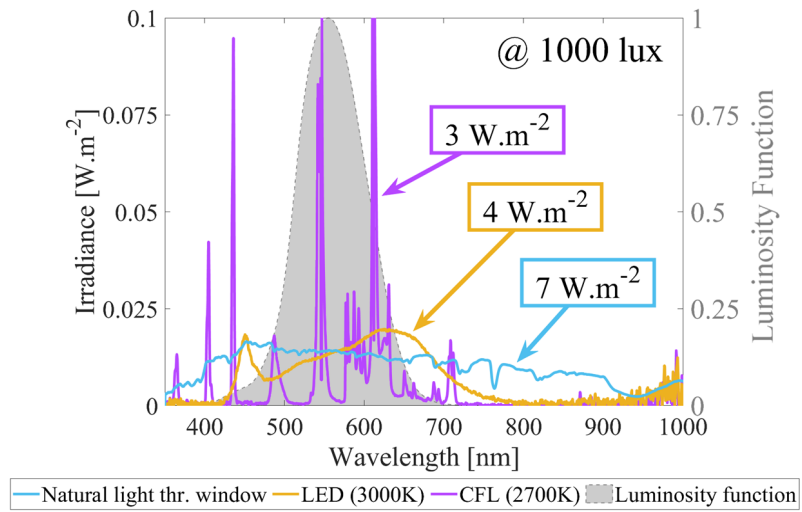

**Fig. 1**. Typical 1000 lux single source indoor environment light spectra: a 3000K LED bulb (in blue), a 2700K CFL bulb (in yellow), and natural light through a window (in red). In green, the luminosity function $Y_\lambda$ used to establish the illuminance level (lux).

In addition, there is a second problem. In the case of real indoor lighting, the incident radiation is a time-varying mixture of multiple natural and artificial direct, reflective, and scattered sources: it has to be taken into account to have a reliable estimation of the photovoltaic cell performance based on real indoor light illumination (Li et al., 2015; Ma et al., 2017; Minnaert and Veelaert, 2014a; Sacco et al., 2013). Finally, the third issue is technological. Indeed, the PV converters are non-ideal devices connected to a non-ideal electrical storage device via a non-ideal power management integrated circuit (PMIC) devices. It means that the electrical energy received by the final consumer device is technology dependent and should be far from the standard theoretical Shockley-Queisser (SQ) limit model predictions (Shockley and Queisser, 1961), whatever the relaxing assumption of the SQ model taken into account (Guillemoles et al., 2019). In the last few years, many researchers have been focused on overcoming these challenges to reach a reliable valuation of the harvestable surrounding low light energy, for example by optimizing the one diode photovoltaic model parameter or new experimental methods for indoor applications (Bader et al., 2019; Fajardo Jaimes and Rangel de Sousa, 2017; Ma et al., 2020; Masoudinejad et al., 2016). But in each of these studies, only the case of controlled mono artificial sources is considered.

In this article, we present a methodology that allows estimating the harvestable energy from any real indoor varying light environment. The first part will present a standard model based on SQ and compare it to some usual real-life PV converters. The second part will present a calculation model based on experimental data focusing mainly on a commercial flexible thin-film Gallium-Arsenide (GaAs) PV solar cell from *Alta Devices Inc.* (Kayes et al., 2011) under several controlled single sources of light. Finally, the third part will present an energy harvesting system prototype based on the GaAs solar cell, working for several days in real varying indoor environments. It is to be noted that this prototype is the micro source of a classic wireless e-ink wifi connected device, with an average power consumption of around 10 mW.

**2. The Shockley-Queisser limit theoretical model adapted to indoor light energy harvesting**

During the last decade, several studies have been conducted to compare the performance of different PV technologies under controlled artificial indoor light single sources (Apostolou et al., 2016; Carvalho and Paulino, 2014; De Rossi et al., 2015; Kasemann et al., 2014; Li et al., 2015; Minnaert and Veelaert, 2014b; Müller et al., 2009). Trying to face the complex task of choosing the best technology to use when dealing with indoor light-harvesting, there is an attractive approach that consists in applying the SQ limit model. This model gives the theoretical maximum conversion efficiency $\eta^{SQ}$ for an ideal semiconductor single-junction solar cell at any theoretical bandgap:

$$\eta^{SQ}(E_g) = \frac{V_{OC}^{SQ} J_{SC}^{SQ} FF^{SQ}}{P_{in}}, \qquad (2)$$

with $P_{in}$ the incident power depending on the spectrum composition, $V_{OC}^{SQ}$ the ideal SQ open-circuit voltage of the PV converter, $J_{SC}^{SQ}$ the ideal SQ short circuit current and $FF^{SQ}$ the ideal SQ fill-factor. In the case of an ideal semiconductor, the short-circuit current equals the photocurrent $J_{ph}^{SQ}$, which is related to the number of incident photons carrying more energy than the bandgap $E_g$:

$$J_{SC}^{SQ} = J_{ph}^{SQ} = q \int_0^{\lambda_g} \Phi_{p,\lambda} \, d\lambda, \qquad (3)$$

Where $\Phi_{p,\lambda}$ is the incident photon flow at each optical wavelength $\lambda$, $q$ the elementary electric charge, and $\lambda_g$ the wavelength corresponding to the considered bandgap $E_g$.

The ideal SQ fill factor $FF^{SQ}$ can be found following the semi-empirical model from Green's approximation (Green and Hall, 1982):

$$FF^{SQ} = \frac{\left[\frac{qV_{OC}^{SQ}}{kT} - ln\left(\frac{qV_{OC}^{SQ}}{kT} + 0.72\right)\right]}{\left(\frac{qV_{OC}^{SQ}}{kT} + 1\right)} \tag{4}$$

with $k$ being the Boltzmann's constant and $T$ the device's temperature. The PV cell open-circuit voltage $V_{OC}$ can be defined by the following equation, in the case of an ideality factor $n$ equal to 1:

$$V_{OC}^{SQ} = \frac{kT}{q} \times ln\left(\frac{J_{ph}^{SQ}}{J_0^{SQ}} + 1\right). \tag{5}$$

Finally, the ideal theoretical dark current density saturation $J_0^{SQ}$ in a single junction is given by (Shockley and Queisser, 1961) and (Müller et al., 2013):

$$J_0^{SQ} = qA\frac{2\pi kT}{h^3 c^2}(E_g + kT)^2 e^{\frac{-E_g}{kT}} \tag{6}$$

with $A$ the active surface area of the PV converter, $h$ the Planck's constant, and $c$ the speed of light in vacuum.

This SQ model is mostly known for having predicted the famous limit efficiency of 33 % for an ideal bandgap around 1.14 eV under the standard (AM 1.5G) sun radiation (Shockley and Queisser, 1961). However, it can also be applied to any kind of spectra like one of the indoor light environments (Müller et al., 2013). For example, Fig. 2(a) shows the result of the SQ model when applied to typical indoor environments, identical to those in Fig. 1, each with an illuminance of 1000 lux. As expected, theoretical efficiencies depend on the spectrum composition and on the fact that the ideal bandgap changes from one single source to another. As a result of this application of the SQ model, for any indoor light, the model gives theoretical efficiency much higher than the solar STC AM1.5G spectrum. In the case of a CFL source, with a theoretical semiconductor with a bandgap around $E_g$ = 1.95 eV, an efficiency as high as $\eta^{SQ}(E_g)$ = 54 % is predicted.

To compare the SQ limit theory to experimental measurement, a commercial source-meter unit (SMU), 2450 from Keithley, has been used to measure $J^{meas}(V)$: the current density-voltage characteristics of different PV converters made of different technologies: hydrogenated amorphous silicon (a-Si:H) solar cell (panel) from *Xiamen Mars Rock Science Technology Co.*, a gallium-arsenide (GaAs) mono-junction flexible PV solar cell from *Alta Devices Inc.* (Kayes et al., 2011), and a polycrystalline silicon (pc-Si) solar cell from *SEEEDSTUDIO*, in controlled indoor single source environments, with different levels of irradiance. The results of the measured fill-factor $FF^{meas}$ extracted from the $J^{meas}(V)$ curves, as well as the $FF^{SQ}$ are plotted in Fig. 2(b).

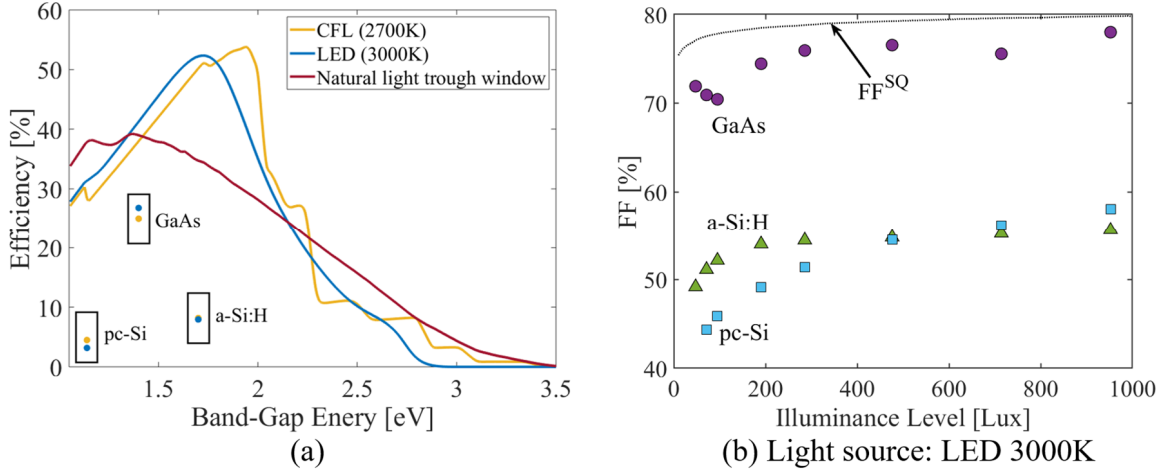

**Fig. 2.** (a) Theoretical maximum conversion efficiency $\eta^{SQ}(E_g)$ of a single ideal p-n junction based on the SQ model for different single sources: CFL, LED, and natural light through a window. Dots are the measured efficiency of different PV converters technology under CFL and LED at around 500 lux. (b) Experimentally measured Fill-factor $FF^{meas}$ of commercial solar cells (GaAs, a-S:H and pc-Si) and their evolution versus LED light intensity. As a comparison, an ideal solar cell $FF^{SQ}$ based on the SQ model is shown in plain black.

The efficiency conversion of the different commercial PV converters technologies is, as expected, different in real conditions from what has been theoretically predicted by the SQ model. It is mostly attributed to several losses, such as shadowing, series and shunt resistance, ideality factor, optical absorption, semiconductor quality, low mobility, small minority carrier lifetime, and the PV converters' fill-factors and thus efficiencies decreasing at low light intensity (Randall and Jacot, 2003). Indeed, $J_{ph}^{SQ}$ varies linearly with the light intensity (3), $V_{OC}^{SQ}$ depends on $J_{ph}^{SQ}$ (5) and $FF^{SQ}$ depends on $V_{OC}^{SQ}$ (4). As an example, the theoretical fill-factor based on the SQ model is shown in Fig .2(b) for a theoretical bandgap corresponding to the standard technologies for indoor application (a-Si:H, pc-Si, and GaAs) and illuminance (from 100 to 1000 lux) with a LED single source. When comparing the experimental results of three commercial PV converters while varying the light intensity of a LED bulb, it can be seen that the fill-factor decreases with lower light intensity, as it is forecast by the theoretical *FF* equation, but with a more dramatic decrease. Indeed, the experimental $FF^{meas}$ of the pc-Si cell strongly decreases when reducing the light intensity, mainly due to their too low shunt resistance (Reich et al., 2009). Thus, as reported previously and exposed in Fig. 2(b), the c-Si technology is not the most suitable technology for indoor applications, and therefore not a good candidate for developing a model based on a one-diode photovoltaic model where the *FF* needs to vary as little as possible. The a-Si:H and GaAs technologies have a more stable *FF* from 100 to 1000 lux, which seems ideal for the simple model presented. Nevertheless, the *FF* of the a-Si:H PV converter is lower than the GaAs cell, and thus its power conversion efficiency (PCE) $\eta^{meas}$ tends to indicate less promising model performance. For indoor applications, especially when dealing with consumer devices with an average consumption of 10 mW, a technology that does not suffer from a low level of light, but also that has a *FF* and a PCE close their ideal value is needed. As seen in Fig. 2(b), only the GaAs cells have demonstrated such good performances, with a measured $FF$ closer to the $FF$ of the Shockley-Queisser model of an ideal solar cell than other considered cells: this is not surprising since III-V solar cells are known to be among the most efficient cells, even under low indoor light conditions (Mathews et al., 2016). This difference in performance is mainly due to the known fact that the carrier mobility is high and the lifetime long in GaAs devices (Kayes et al., 2011), that the series resistance is low enough, and the shunt resistance high enough to avoid deviation of the FF, even in a low light environment. In conclusion, the fact that the experimental $FF^{meas}$ is almost constant under an illuminance ranging from 100 to 1000 lux considerably simplifies the theoretical calculation of the output power of a GaAs PV cell from any spectrum, by safely applying the superposition model, even at a low light intensity.

The next paragraph will describe a methodology based on the superposition model to predict the extractable energy from the GaAs thin-film solar cell, whatever the real indoor light source environment.

**3. A calculation model of indoor light energy harvesting based on a one-diode photovoltaic model**

The method is based on the combination of the results from three measurements: i) the light environment spectrum (which is fluctuating in time), ii) the external quantum efficiency ($EQE$) of the PV cell (unchanging intrinsic values) and iii) the measurement of the current density-voltage $J_{dark}^{meas}(V)$ characteristic in the dark (unchanging intrinsic values). This simple methodology allows us to consider the primary deviations from the SQ model: a real $EQE$ instead of an ideal one over the whole range of the spectra, and a real $J_{dark}^{meas}(V)$ curve, which takes into account the real "shape"($FF$) of the current density -voltage $J(V)$ characteristic, including the parasitic series $R_s$ and shunt $R_{sh}$ resistances as well as the ideality factor $n$, which can vary from 1 to 2. The spectral response was measured with a custom-built setup composed of a Xenon lamp, a monochromator equipped with two diffraction gratings, a filter wheel to remove the higher diffraction orders of radiation, and a lock-in amplifier. The measurements were calibrated with Si and Ge photodiodes (Thorlabs FDS100-CAL and FDG03-CAL, respectively) to cover the whole wavelength range of interest. This characterization does not consider the impact of the solid angle of the radiation. It studies the characteristics of the PV converter with a photon flow that is normal to its surface. For each wavelength band produced by the source, the current generated by the characterized PV cell is measured. A known cell is used as a current-generation reference to establish the EQE of the characterized PV cell by comparison.

In the developed model, the measured $EQE$ is added in (3) to consider the deviation from the SQ model due to the real external quantum efficiency of the PV cell. The resulting short-circuit current density $J_{SC}^{QE}$ is thus an estimation based on two measurements: the $EQE$ which is fixed for each PV converter and the spectra $\Phi_{p,\lambda}$ fluctuating in time:

$$J_{SC}^{QE} = q \int_0^{\lambda_g} \Phi_{p,\lambda} \times EQE \, d\lambda \tag{7}$$

Now that the spectral response $EQE$ is not ideal anymore but is measured from a real solar cell, the equation (7) gives an estimated value of the short circuit density current $J_{SC}^{QE}$ lower than in the ideal case based on the SQ model ($J_{SC}^{SQ}$). This value depends on the real measured spectra at any given time. Then, the measured dark current density-voltage curves $J_{dark}^{meas}(V)$ of the PV converters can be collected by the SMU 2450 from Keithley, and using the superposition principle (Lindholm et al., 1979; Tarr and Pulfrey, 1980) or more commonly known as the one-diode photovoltaic model, it is possible to deduce the estimated current density-voltage $J^{mod}(V)$ model at any given time of the fluctuating spectra:

$$J^{mod}(V) = J_{dark}^{meas}(V) + J_{SC}^{QE} \tag{8}$$

As an example, Fig. 3(a) shows the LED spectra of a 300 lux LED 3000K bulb as well as the measured $EQE$ curve of a GaAs solar cell and the current generation of this cell resulting from its absorption of the photon flux. In Fig. 3(b), based on the previous $EQE$ curve and the LED spectra and the $J_{dark}^{meas}(V)$ curve, the estimated $J^{mod}(V)$ is shown as well as the measured $J^{meas}(V)$. In this first example, the current density curve $J^{mod}(V)$, estimated from the developed model, is in good agreement with the experimentally measured $J^{meas}(V)$. In this example, the extracted maximal power values $P_{max}^{mod}$ and $P_{max}^{meas}$ from the model $J^{mod}(V)$ and measured $J^{meas}(V)$, are 30 µW/cm² and 31 µW/cm², respectively, with a difference of about 3 %.

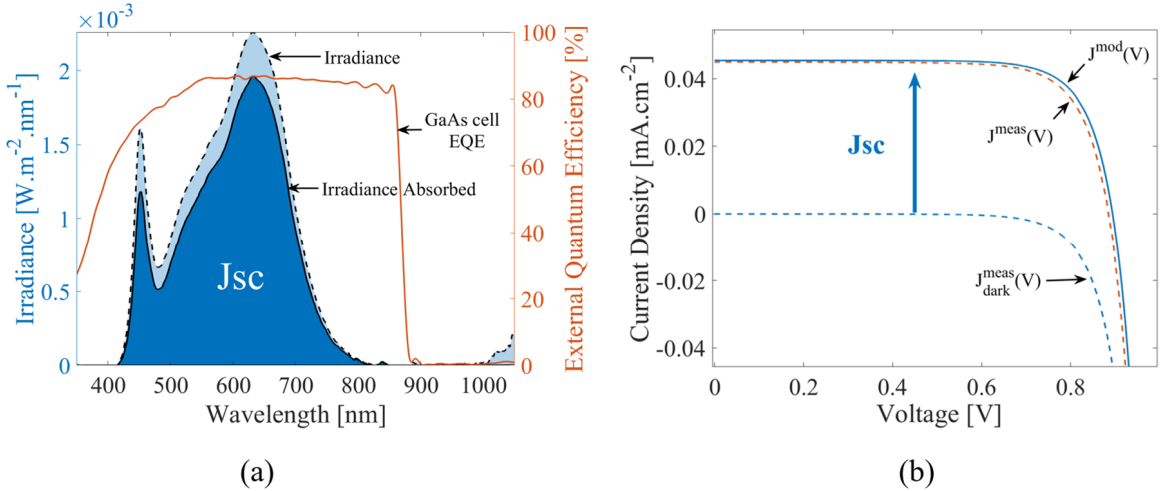

(a)          (b)

**Fig.3.** (a) spectra of a 300 lux LED 3000K bulb and the same spectra modulated by the measured external quantum efficiency of a GaAs solar cell. (b) the $J_{dark}^{meas}(V)$ curve of the GaAs solar cell in blue dash used to calculate the expected $J^{mod}(V)$ based on (8) in orange dash, compared to the experimentally measured $J^{meas}(V)$.

We have used this method with the same GaAs solar cell under single controlled sources light environments, for different sources (CFL and LED) at many different levels of indoor irradiance (from 200 lux up to 1000 lux), to verify the reliability of the one-diode photovoltaic model. The characteristics resulting from the model, $J^{mod}(V)$ and the $P^{mod}(V)$, are shown in Fig. 4 and compared to their experimental equivalents, $J^{meas}(V)$ and the $P^{meas}(V)$. The deviation between model results and the experimental measurements is noticeable but minor. With an average error of less than 3 % and never exceeding 5 %, considering the numerous experimental bias possible, these results seem sufficiently accurate to be useful for the calculation of a harvestable power approximation in various low irradiance indoor environments.

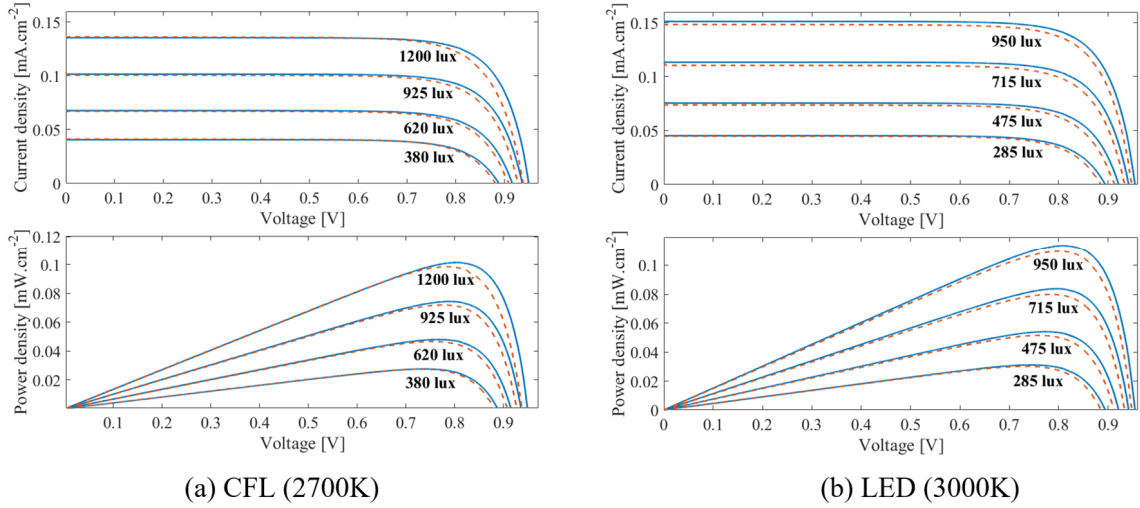

(a) CFL (2700K)  (b) LED (3000K)

**Fig. 4** Comparison between the calculated $J^{mod}$-V and the $P^{mod}$-V (plain blue lines) from the one-diode photovoltaic model and the corresponding experimental characteristics $J^{meas}$-V and the $P^{meas}$-V (dotted orange lines) of a GaAs solar cell for different illuminance under (a) a CFL light source and (b) a LED light source.

As stated previously in this paper, the lack of standards to rely on makes studying in the field of indoor PV more complex. Consequently, it is necessary to find a method to evaluate the produced energy based on the analysis of a fluctuating mix of natural and artificial light environments. The next paragraph will demonstrate the reliability of the simple one-diode photovoltaic model described beforehand to calculate the harvestable energy in real-life indoor conditions, even for several days.

## 4. The model applied in real conditions for several days

The ultimate purpose of this study is to establish a model to calculate, as precisely as possible, the level of harvestable energy in a real indoor environment for an extended period. Therefore, it is necessary to test the model under real-life conditions, observing the light environment over time to know the variations in its composition and to compare the calculated results to experimental results. A complete energy harvesting prototype, shown in Fig. 5, has been developed to validate the developed model experimentally in real-life environments. This prototype is based on two GaAs thin-film solar cells providing electrical energy to the energy storage device (here a Lithium-Polymer battery) of a consumer device like an e-ink connected device. To extract the maximum power from the solar cells, a very low consumption commercial power management integrated circuit (PMIC) from *e-PEAS* has been placed between the PV cells and the LiPo battery.

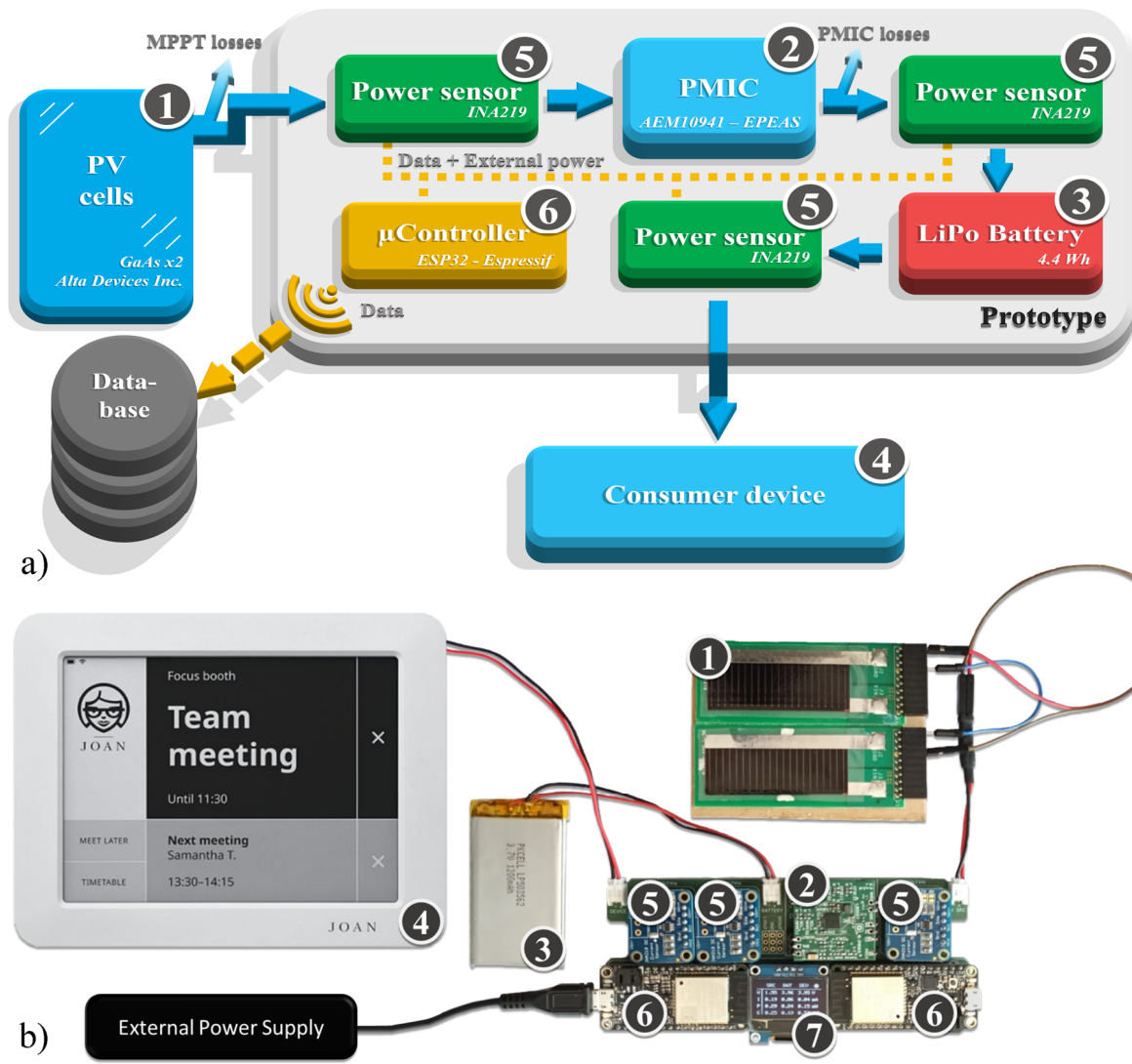

**Fig. 5.** a) Schematic of the energy harvesting prototype (PV cells + PMIC + LiPo battery) that has been instrumented by three INA 219 electrical power sensors and two ESP32 microchips to collect the data and send them to a database. Let's note there is a specific external power source that allows the acquisition and transmission of data without affecting the performance of the energy harvester.b) a picture of the described prototype integrating GaAs solar cells (1), a PMIC : an e-peas 'EVK10941M' Mini Evaluation Board based on the 'AEM10941' Solar Energy Harvesting IC (2); a LiPo electrical storage device (3), the e-ink communicating device to power (4), the INA219 power sensors (5), the ESP32 low cost low power microcontroller chips (6) and an OLED screen (7).

The used PMIC is based on a max power point tracking (MPPT) algorithm called fractional open-circuit voltage (FOCV). This method relies on getting a sample of the $V_{oc}$ and then applies and holds a fraction of its value to the PV cells or module (Motahhir et al., 2020). Besides not being the most performant in casual energy harvesting situations, such an MPPT algorithm has demonstrated its use of relevance in low light environments (Weddell et al., 2012). Even though this PMIC is adapted for low energy harvesting applications, the losses are not negligible. To include these losses in our model, and to be able to measure the "real" power harvested by the prototype at any time into the LiPo battery, two voltage and current sensing chips INA219 from *Texas Instrument* have been inserted between the PV cells and the PMIC and between the PMIC and the LiPo. Regarding the classical Lithium Polymer (LiPo) battery whom internal resistance is usually less than 100mΩ, it has no impact in term of losses when considering the low current level in our application. In addition, a third INA219 has been inserted between the LiPo battery and the consumer device to monitor its consumption. A set of two microcontrollers associated with these power sensors allows the acquisition of the power generation/consumption data from the three INA 219, and their recording on a database in the cloud. An external power source is used to power the sensors and the microcontrollers.

The e-peas PMIC Mini Evaluation Board EVK10941M chosen for the prototype uses the fractional open-circuit voltage (FOCV) algorithm. To extract the maximum power from the PV cells, this PMIC algorithm can be configured to apply 70%, 75%, 85%, or 90% of their $V_{oc}$. In our application, by choosing the fraction of 85%, the MPPT losses due to the optimal fraction of $V_{oc}$ not being exactly equal to 85%, are evaluated to be lower than 1% over the whole range of illumination. Regarding the PMIC losses induced by the classical losses of a buck-boost converter, they can vary from 10% to 20% depending on the level of illumination. As can be seen in Fig.6, at low light (and low $V_{SRC}$ et low $I_{SRC}$), the efficiency of the PMIC is only 80 % while at moderate to high illumination the efficiency is about 90%. From the plot of Fig.6, a regressive equation has been added to our model to include the PMIC efficiency $\eta^{PMIC}$.

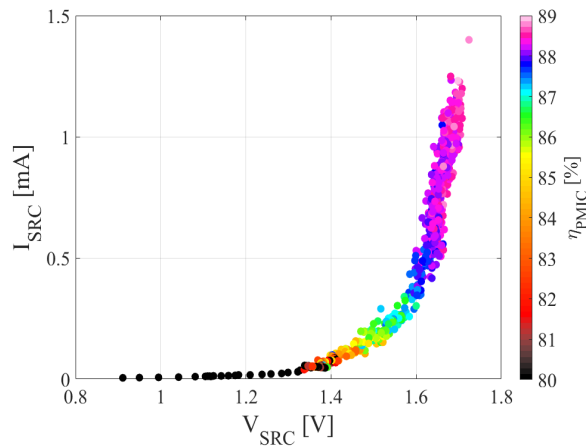

**Fig. 6.** Experimental PMIC efficiency $\eta^{PMIC}$ depending on ligh intensity. These curve depends on the experimental set-up , in our case two GaAs solar cell in series , an epeas based PMIC and a single cell LiPo.

Finally, the purpose of the experiment is to compare the harvestable energy calculated by the model and the real harvested energy measured by the power sensor, the power measurements must be carried out simultaneously with the spectrometer analysis of the light environment to which the prototype is exposed. This experiment has been carried out over 21 days. The results of 2 typical days (sunny and cloudy days) are shown in Fig.7. While the orange dashed line is the experimental harvested power stored into the Lithium Polymer battery measured directly by an INA219, the blue line represents the model calculation results based on one-minute spectra measurement intervals. A qualitative agreement between the model calculation of power and energy harvesting and the real usable power generated and energy transferred to the prototype battery can be observed. The mean absolute percentage error (MAPE) between the model curve and the experimental power curve of each of the 21 days is lower than 6%. These very good results are comforting the fact that our model and methods are reliable enough to be an interesting tool for researchers and engineers conceiving indoor micro source elements for IoT applications or consumer devices.

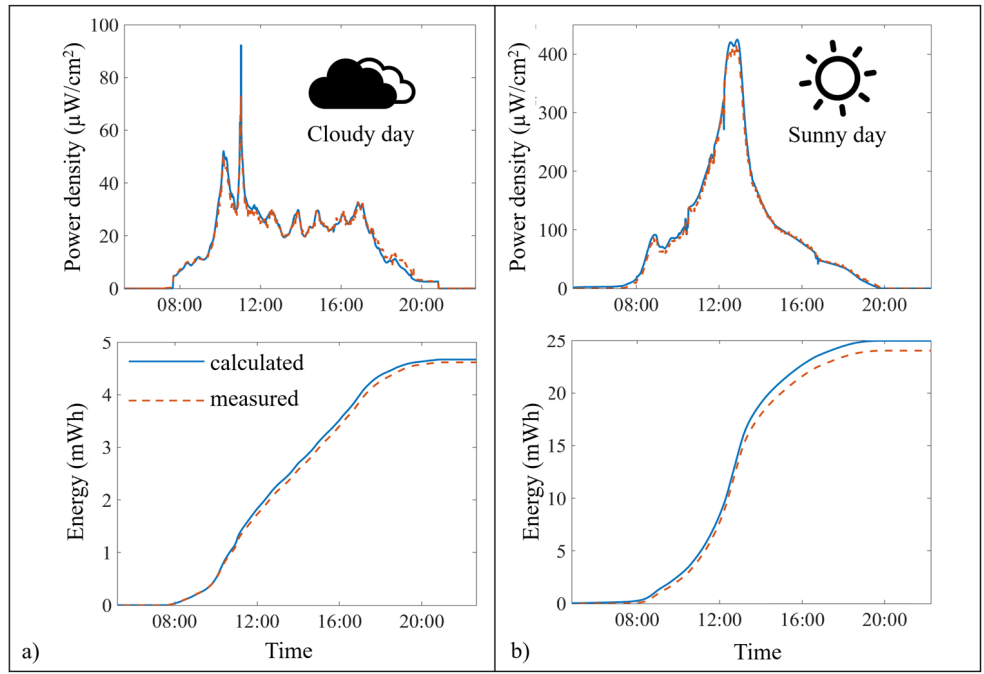

**Fig. 7.** model's power density and energy calculations (plain blue) and experimentally measured power-density and energy (dash orange) generated from the GaAs solar cell in a real-life indoor environment harvested into the Lithium Polymer storage device for a) a cloudy day and b) a sunny day.

The 6% of error are potentially due to several factors: i) the MPPT losses ii) the LiPo battery losses, iii) the reliability of the low-cost INA219, iv) the reliability of our model which has been taken as simple as possible without taking into account potential variation of $R_{sh}$ and $R_s$ with light intensity, v) the small effect of temperature in indoor Environnement, vi) and of course the experimental setup.

These results obtained after 21 days of observation of the office's real lighting environment show that the model makes it possible to determine the level of harvestable energy per surface unit quite accurately, even if there is still room for improvement. It is then possible to determine, based on these 21 days of observation, the harvesting surface area or the number of PV cells required in this environment to compensate for the energy needs of an electronic device fully. For example, a classical wireless e-ink dashboard, as the one used in this study, depending on its size and the significance of its sleeping mode, consumes about 10 mW on average. In such a case, a preliminary extrapolation shows that 14x14 cm² of similar solar cells would be sufficient to supply it enough power to make it energetically autonomous.

## 5. Conclusion

In summary, the simple model presented applying the superposition principle, based on the combination of the measured optical EQE and electrical $J_{dark}^{meas}$-V of GaAs PV cell in the dark, combined with the spectral measurements of real-life indoor environments, has demonstrated the ability to calculate the potentially harvestable energy in the environment that is being studied. Using an energy-harvesting prototype installed in the environment studied by the spectrometer feeding the developed model allowed to confront results from the model to actual power and energy harvest. The confrontation result corroborates the ability of the proposed model to calculate the harvestable energy with an error lower than 6 % for the 21 days of the test. In a more practical sense, this paper shows that in real-life indoor environments, a module based on PV cells similar to *Alta Devices Inc.* GaAs PV solar cells and smaller than a third of a sheet of paper (A4 format) is enough to make devices consuming about 10 mW.

We believe that the method presented in this article will be helpful to engineers and researchers attempting to overcome the various challenges toward the accurate design of energy harvesting devices in the future.

**Declaration of Competing Interest**

The authors declare no conflict of interest.

**Acknowledgments**

This work has been funded by the French national association for research and technology (ANRT) [grant CIFRE number 2017/0331].